\documentclass[aps,prb,reprint,showpacs,twoside,preprintnumbers]{revtex4-1}
\usepackage{hyperref}
\usepackage{graphicx}
\usepackage[utf8x]{inputenc}

\begin{document}
\preprint{The original publication is available at prb.aps.org: Phys. Rev. B 83,
235120 (2011); DOI:10.1103/PhysRevB.83.235120}
\title{Electronic transport coefficients from ab initio simulations and
application to dense liquid hydrogen  }
\author{Bastian Holst}
\author{Martin French}
\author{Ronald Redmer}
\affiliation{Institut f\"ur Physik, Universit\"at Rostock, D-18051 Rostock, Germany}
\date{\today}

\begin{abstract}
Using Kubo's linear response theory, we derive expressions for the
frequency-dependent electrical conductivity (Kubo-Greenwood formula),
thermopower, and thermal conductivity in a strongly correlated electron system.
These are evaluated within {\it ab initio} molecular dynamics simulations in
order to study the thermoelectric transport coefficients in dense liquid
hydrogen, especially near the nonmetal-to-metal transition region. We also
observe significant deviations from the widely used Wiedemann-Franz law which is
strictly valid only for degenerate systems and give an estimate for its valid
scope of application towards lower densities.
\end{abstract}



\maketitle

\section{Introduction}

An accurate description of dense hydrogen is of fundamental interest and has
wide applications in astrophysics. Prominent examples are models of planetary
interiors which consist predominantly of hydrogen and helium as in the case of
Jupiter-like planets.~\cite{Guillot1999, Nettelmann2008, Fortney2010} Special
attention has been paid on the high pressure phase diagram of hydrogen and its
isotopes. Major problems in this context are the slope of the melting
line~\cite{Bonev2004a, Gregoryanz2003, Deemyad2008, Eremets2009} and the
transition pressure to solid metallic hydrogen,~\cite{Loubeyre2002, Pickard2007}
which is expected near 4~Mbar~\cite{Stadele2000} at $T=0$~K. Such extreme
conditions are experimentally still not feasible, yet. 

Interestingly, metallic hydrogen has first been verified in the liquid at about
1.4~Mbar and few thousand Kelvin.~\cite{Weir1996, Nellis1999} The problem
whether or not this transition is of first order has been discussed over
decades. While numerous models within the chemical picture predict almost
invariably a pronounced first-order phase transition with a critical point at 
about 10000-15000~K and 0.5~Mbar,~\cite{Ebeling1985, Saumon1989, Schlanges1995, 
Reinholz1995, Beule1999, Holst2007} only few {\it ab initio} simulations indicate 
such a behavior.~\cite{Filinov2001, Scandolo2003, Jakob2007, Morales2010a,
Lorenzen2010}  A first-order liquid-liquid phase transition with a critical
point near 2000~K and 120~GPa has been predicted from both quantum Monte Carlo
(QMC) and finite-temperature density functional theory molecular dynamics
(FT-DFT-MD) simulations.~\cite{Morales2010a} These results were confirmed almost
simultaneously by extensive FT-DFT-MD simulations~\cite{Lorenzen2010} which
predict a critical point at about 1400~K and 132~GPa, i.e.\ at a somewhat lower
temperature. A detailed analysis of the changes in the structural and electronic
properties with density and temperature clearly shows that the nonmetal-to-metal
transition in dense liquid hydrogen drives this first-order phase transition. 

The transport coefficients yield valuable information on the state of the 
strongly correlated liquid. For instance, the conductivity changes drastically 
along the nonmetal-to-metal transition over many orders of magnitude which could 
already be verified experimentally.~\cite{Weir1996, Nellis1999} These measures 
have the potential to accurately characterize subtle changes of the electronic 
structure with density and temperature and, especially, alongside the 
liquid-liquid phase transition in hydrogen. We therefore calculate the complete
set of thermoelectric transport coefficients (electrical and thermal
conductivity, thermopower) via FT-DFT-MD simulations for a wide range of
densities ($\rho=(0.05-20)$~g/cm$^3$) and temperatures
($T=(1-50)\times10^3$~K) and pay special attention to the
nonmetal-to-metal transition region. The focus of our work is below  20~g/cm$^3$
since higher densities are of primary importance for the physics of inertial
confinement fusion and were studied previously.~\cite{Recoules2009} 

The transport coefficients as well as their changes with density and temperature
are again important for applications in astrophysics. For instance, a boundary
between a nonconducting outer and a metallic inner envelope is usually assumed
in interior models of gas giants such as Jupiter; its location should in
principle be determined from the nonmetal-to-metal transition in hydrogen but is
usually a free parameter.~\cite{Saumon1995} Furthermore, it has been shown that
demixing of helium from hydrogen occurs in H-He mixtures at megabar pressures
due to the nonmetal-to-metal transition in the hydrogen
subsystem.~\cite{Lorenzen2009, Morales2009} The treatment of this effect is
essential in order to explain the excess luminosity of Saturn and its
age.~\cite{Fortney2004} Knowledge on the transport coefficients of dense liquid
hydrogen is furthermore required for dynamo simulations of planetary magnetic
fields.~\cite{Wicht2010} The dynamo is driven by convection of conducting
material deep in the interior so that the electrical and thermal conductivity
are important input quantities. Another interesting problem is the formation of
giant planets out of a protoplanetary disc. The conductivities change by many
orders of magnitude during the accretion process as a consequence of the density
and temperature increase which could affect the radiation hydrodynamics of the
collapsing disc and the luminosity of the young protoplanet strongly.

There are a number of theoretical models that predict the electrical and thermal
conductivity with different assumptions about the electronic and ionic structure
and their mutual interaction.~\cite{Spitzer1953, Minoo1976, BRD1982, Lee1984,
Hoehne1984, Ichimaru1985, Rinker1988, Kitamura1995, Reinholz1995, Redmer1997}
Especially in a strongly coupled system it is difficult to calculate the
transport properties accurately because the ionization degree and the effective
two-particle scattering cross sections are not well defined. Strong ion-ion
correlations, the dynamic nature of screening and exchange effects in the
electron system as well as quantum effects such as Pauli blocking require a
consistent quantum-statistical approach. Therefore, we calculate the transport
coefficients within FT-DFT-MD simulations. This method has demonstrated its
capacity to provide accurate data for the strongly correlated quantum
regime.~\cite{Desjarlais2002, Recoules2005, Holst2008, Caillabet2011}

For simple metals the relation between electrical and thermal conductivity is
described by the famous Wiedemann-Franz law~\cite{Wiedemann1853} using a fixed 
value, the Lorenz number $L_0$, as proportionality constant. The benefit from 
this relation is the ability to obtain the thermal conductivity from the 
electrical conductivity very easily. For this procedure the Lorenz number
$L(T,\rho)$ has to be known for a given density $\rho$ and temperature $T$ of
the system. By calculating the Lorenz number {\it ab initio} the region where
the original Wiedemann-Franz law is valid can be identified.

Our paper is organized as follows. We outline the FT-DFT-MD method in
Sec.~\ref{sec:theory}. A generalization of the Kubo-Greenwood formula to
calculate the Onsager coefficients $L_{ik}$ is given in Sec.~\ref{sec:Lik}. The
results for the electrical and thermal conductivity, the thermopower, and the
Lorenz number in dense liquid hydrogen are presented in Sec.~\ref{sec:res} for a
wide range of densities and temperatures. In particular, the behavior predicted
for the nonmetal-to-metal transition region is discussed. Conclusions are given
in Sec.~\ref{sec:concl}.

\section{Theoretical method}\label{sec:theory}

We use the FT-DFT-MD framework which combines classical molecular dynamics
simulations for the ions with a quantum treatment of the electrons based on
FT-DFT~\cite{Mermin1965, Wentzcovitch1992, Weinert1992} which is implemented in
the VASP 5.2 program package.\cite{Kresse1993, Kresse1994, Kresse1996} In the
FT-DFT the Coulomb interactions between the electrons with the ions are treated
using the projector-augmented wave method~\cite{Blochl1994, Kresse1999} at
densities below 9~g/cm$^3$ with a converged energy cutoff of 800~eV. At higher
densities it was necessary to perform all calculations with the Coulomb
potential which required a substantial higher cutoff of 3000~eV.

The FT-DFT algorithm is used to derive the forces that act on the ions via the
Hellmann-Feynman theorem at each MD step. This procedure is repeatedly performed
in a cubic simulation box with periodic boundary conditions for several thousand
MD time steps of $0.1$~fs to 1~fs duration so that the total simulation time 
amounts up to 10~ps. The ion temperature is controlled with a Nos\'e
thermostat.~\cite{Nose1984} 

Convergence was checked with respect to the particle numbers, which vary between
64 and 512 atoms depending on the density, the $\mathbf{k}$-point sets used for
the evaluation of the Brillouin zone, and the energy cutoff for the plane wave
basis set. For the simulations we chose the Baldereschi mean value 
point~\cite{Baldereschi1973} which proved to yield well converged simulation
runs.~\cite{Holst2008, Lorenzen2010} Several test calculations showed that higher 
efforts are necessary only in the vicinity of the phase transition.~\cite{Lorenzen2010} 

The electronic transport coefficients are subsequently calculated by
evaluating the respective transport formulas, see section below.~\cite{VASPOPTIC} 
This is done for 10 to 20 ion configurations from the equilibrated MD simulation 
using Monkhorst-Pack $\mathbf{k}$-point meshes~\cite{Monkhorst1976} of 
$3\times 3\times 3$ to $6\times 6\times 6$ to reach the convergence. 

The exchange-correlation functional is the most critical input in FT-DFT
calculations. Here we use the approximation of Perdew, Burke, and
Ernzerhof~\cite{Perdew1996} which is numerically efficient. This functional has 
been chosen in similar studies~\cite{Recoules2009, Boates2011, Horner2010} 
and reasonable results are in general expected in the metallic and high-temperature 
plasma regime.~\cite{Faleev2006, French2010} In the semiconducting region the obtained
conductivities may be overestimated by using this functional but will yet be useful 
for many practical applications.

Note that the ionic contribution to the transport coefficients is out of
the scope of this work and is therefore consistently neglected.

\section{Transport properties and Onsager coefficients}\label{sec:Lik}

In linear response theory (LRT) the response of an isotropic single component
system of charge carriers (electrons with charge $q=-e$ in our case) to an
electric field \textbf{E} and a temperature gradient $\nabla T$ is expressed by
the electric current  $\mathbf{J}_e$ and the heat current $\mathbf{J}_q$
through:
\begin{eqnarray}
\label{eq:elec_cur}
 \langle \mathbf{J}_e \rangle &=&
\frac{1}{q}\left(qL_{11}\textbf{E}+\frac{L_{12}\nabla
T}{T}\right)\quad\mathrm{,}\\
\label{eq:heat_cur}
\langle \mathbf{J}_q \rangle &=&
\frac{1}{q}\left(qL_{21}\textbf{E}+\frac{L_{22}\nabla
T}{T}\right)\quad\mathrm{.}
\end{eqnarray}
The electrical conductivity $\sigma$, thermal conductivity $\lambda$, and
thermopower $\alpha$ are then given by the Onsager coefficients $L_{mn}$ in the
following way:
\begin{equation}
 \sigma=L_{11} \mathrm{,}\quad\lambda=\frac{1}{T}
\left(L_{22}-\frac{L^2_{12}}{L_{11}}\right) \mathrm{,}\quad
\alpha=\frac{L_{12}}{TL_{11}}\quad\mathrm{.} \label{eq:trans_coeff}
\end{equation}
The Lorenz number is the ratio between electrical and thermal conductivity
divided by the temperature,
\begin{equation} \label{LorenzN}
 L=\frac{e^2}{k_B^2T}\frac{\lambda}{\sigma} \quad\mathbf{,}
\end{equation}
and is, according to the Wiedemann-Franz law,\cite{Wiedemann1853} constant in
the limit of high density, where it can be calculated as $L_0=\pi^2/3$
by means of the Sommerfeld expansion.\cite{Ashcroft1976}

Within the framework of Kubo's quantum-statistical LRT, which is described for
instance in Ref.~\protect\onlinecite{Christoph1985, Zubarev1997}, the following
expressions are obtained for the frequency-dependent
Onsager coefficients $L_{mn}(\omega)$:
\begin{equation}
L_{mn}\left(\omega \right)=\frac{1}{3V}\langle
\hat{\mathbf{J}}_m\left(t-i\hbar\tau\right);\hat{\mathbf{J}}_n\rangle_{
\omega+i\varepsilon }\quad\mathrm{.}
\end{equation}
The current-current correlation functions are given as: 
\begin{eqnarray}
\langle
\hat{\mathbf{J}}_m\left(t-i\hbar\tau\right);\hat{\mathbf{J}}_n
\rangle_{\omega+i\varepsilon}
=\lim_{\varepsilon \to 0}\int\limits_{0}^{\infty} \text{d} t \,
e^{i\left(\omega+i\varepsilon\right)t}\nonumber\\\times
\int\limits_0^\beta \text{d} \tau \,
\text{Tr}\left\{\hat{\varrho}_0\hat{\mathbf{J}}_m
\left(t-i\hbar\tau\right)\cdot\hat{ \mathbf{J}}_n \right\}\quad\mathrm{,}
\label{eq:cccf}
\end{eqnarray}
where $\beta=(k_BT)^{-1} $ is the inverse thermal energy.
The limit  $\varepsilon \to 0$ has to be taken after the calculation of the
thermodynamic limit, which is done by evaluating the trace. 
The statistical operator of the equilibrium $\hat{\varrho}_0$ contains the
Kohn-Sham Hamilton operator $\hat{H}_\mathrm{KS}$. 

The time-dependent current operators within the
Heisenberg picture are defined as:
\begin{equation}
\hat{\mathbf{J}}_m\left(t-i\hbar\tau\right)=e^{\frac{i}{\hbar}
\left(t-i\hbar\tau\right)\hat{H}_\mathrm{KS}}\hat{\mathbf{J}}_me^{-\frac{i}{
\hbar }
\left(t-i\hbar\tau\right)\hat{H}_\mathrm{KS}} \quad\mathrm{.}
\end{equation}
The electric current operator $\hat{\mathbf{J}}_e=\hat{\mathbf{J}}_1$ and the
heat current operator $\hat{\mathbf{J}}_q=\hat{\mathbf{J}}_2$ read in second
quantization of spin degenerate Bloch states:
\begin{equation}
\label{eq:CurOpI}
\hat{\mathbf{J}}_m= \sum_{\mathbf{kk}'\nu\nu'} \langle
\mathbf{k}\nu
|\hat{\mathbf{j}}_m|
\mathbf{k'}\nu' \rangle \hat{a}_{\mathbf{k}\nu}^{\dagger}
\hat{a}_{\mathbf{k'}\nu'}^{\,_{\,}} \quad \mathrm{.}
\end{equation}
Here $\mathbf{k}$ is the wave number and $\nu$ the band index. The electric and
heat current operators are given as
\begin{eqnarray}
\hat{\mathbf{j}}_1&=&\frac{q}{m_e}\hat{\mathbf{p}}\quad\mathrm{,}\\
\hat{\mathbf{j}}_2&=&\frac{1}{m_e}
\frac{\hat{H}\hat{\mathbf{p}}+\hat{\mathbf{p}}\hat{H}}{2}
-h_e\hat{\mathbf{ p}}\quad\mathrm{,} \label{ws}
\end{eqnarray}
where $h_e$ is the enthalpy per electron, see Ref.~\protect\onlinecite{deGroot1984}
for further information on the definition of the currents.

The eigenvalues of the Hamiltonian can be evaluated so that Eq.\ (\ref{eq:CurOpI}) 
can be simplified,
\begin{equation}
\label{eq:CurOpII}
\hat{\mathbf{J}}_m=\frac{q^{2-m}}{m_e} \sum_{\mathbf{kk}'\nu\nu'} \langle
\mathbf{k}\nu
|\hat{\mathbf{p}}|\mathbf{k'}\nu' \rangle
\epsilon_{\mathbf{k}\nu\mathbf{k}'\nu'}^{m-1}
\hat{a}_{\mathbf{k}\nu}^{\dagger}
\hat{a}_{\mathbf{k'}\nu'}^{\,_{\,}} \quad \mathrm{,}
\end{equation}
where
\begin{equation}
\epsilon_{\mathbf{k}\nu\mathbf{k}'\nu'}=\left(\frac{E_{\mathbf{k}\nu}+E_{\mathbf
{k}'\nu'}}{2}-h_e\right) 
\end{equation} 
is used. After inserting Eq.\ (\ref{eq:CurOpII}) into Eq.\ (\ref{eq:cccf}) and
some operator algebra the trace can be evaluated according to Wick's theorem as:
\begin{eqnarray}
&\text{Tr}&\left\{ \hat{\varrho}_{0}  \hat{a}_{\mathbf{k}\nu}^{\dagger}
\hat{a}_{\mathbf{k'}\nu'}^{\,_{\,}} \hat{a}_{\mathbf{p}\mu}^{\dagger}
\hat{a}_{\mathbf{p'}\mu'}^{\,_{\,}} \right\}=\nonumber\\
&&f_{\mathbf{k}\nu} \delta_{\mathbf{k},\mathbf{k'}}
\,\delta_{\nu,\nu'}f_{\mathbf{p}\mu} \delta_{\mathbf{p},\mathbf{p'}}
\,\delta_{\mu,\mu'}\nonumber\\
&&+f_{\mathbf{k}\nu} \delta_{\mathbf{k},\mathbf{p'}}
\,\delta_{\nu,\mu'}\left(1-
f_{\mathbf{p}\mu}\right)\delta_{\mathbf{p},\mathbf{k'}}
\,\delta_{\mu,\nu'} \quad \mathrm{.}
\label{eq:Wick}
\end{eqnarray}
The first term vanishes and the Fermi functions are defined as
$f_{\mathbf{k}\nu}=(e^{\beta\left(E_{\mathbf{k}\nu}-\mu_e\right)}+1)^{-1}$.
Altogether the trace reads
\begin{eqnarray}
&&\text{Tr}\left\{\hat{\varrho}_0\hat{\mathbf{J}}_m
\left(t-i\hbar\tau\right)\cdot\hat{ \mathbf{J}}_n \right\}
=\frac{q^{4-m-n}}{m_e^2}
\sum_{\mathbf{kp}\nu\mu}
e^{\frac{i}{\hbar}\left(t-i\hbar\tau\right)\Delta E}\nonumber
\\
&&\nonumber
\\
&&\times
f_{\mathbf{k}\nu} \left(1- f_{\mathbf{p}\mu}\right)
\langle\mathbf{k}\nu |\hat{\mathbf{p}}|\mathbf{p}\mu \rangle\cdot
\langle \mathbf{p}\mu |\hat{\mathbf{p}}|\mathbf{k}\nu \rangle
\epsilon_{\mathbf{k}\nu\mathbf{p}\mu}^{m-1}
\epsilon_{\mathbf{p}\mu\mathbf{k}\nu}^{n-1}
\;\mathrm{,}
\end{eqnarray}
with $\Delta E=E_{\mathbf{k}\nu}-E_{\mathbf{p}\mu}$.
Now the $\tau$ integration can be performed:
\begin{equation}
\int\limits_{0}^{\beta} \text{d} \tau \, e^{\tau
\Delta E} = \frac {e^{\beta\Delta E}-1}{\Delta E} 
\quad\mathrm{.}
\end{equation}
In the same way the second integral can be solved:
\begin{eqnarray}
\lim_{\varepsilon\to0} 
\int\limits_{0}^{\infty} \text{d} t \,&& e^{\left(-\varepsilon + i\omega+
\frac{i}{\hbar} \Delta E\right) t} 
= \nonumber\\ 
\hbar \pi \delta\left(\Delta
E+\hbar\omega\right) 
&&+ i \hbar \, \mathcal{P} \left(\frac{1}{\Delta
E+\hbar\omega} \right)
\quad \mathrm{.} 
\end{eqnarray}
Only the real part of this equation is considered here, since the imaginary
part can be calculated more easily with a Kramers-Kronig relation. The Onsager
coefficients now read:
\begin{eqnarray}
L_{mn}&=&\frac{\hbar\pi q^{4-m-n}}{3Vm_e^2}
\sum_{\mathbf{kp}\nu\mu} \frac{e^{\beta\Delta E}-1}{\Delta E} f_{\mathbf{k}\nu}
\left(1-f_{\mathbf{p}\mu}\right) 
\\\nonumber
&\times&
\langle \mathbf{k}\nu |\hat{\mathbf{p}}|\mathbf{p}\mu \rangle\cdot
\langle \mathbf{p}\mu |\hat{\mathbf{p}}|\mathbf{k}\nu \rangle
\delta\!\left(\Delta E+\hbar\omega\right)
\epsilon_{\mathbf{k}\nu\mathbf{p}\mu}^{m-1}
\epsilon_{\mathbf{p}\mu\mathbf{k}\nu}^{n-1}
\mathrm{.}
\end{eqnarray}
In position representation the matrix elements have the following form:
\begin{equation}
\langle \mathbf{k}\nu |\hat{\mathbf{p}}|\mathbf{p}\mu
\rangle = \delta_{\mathbf{p},\mathbf{k}} \!\left( \hbar\mathbf{k}
\,\delta_{\nu,\mu}+\frac{1}{V} \!\int\limits_{V} \!\text{d}^3
\mathbf{r}\, u_{\mathbf{k}\nu}^\ast\left(\mathbf{r} \right) \hat{\mathbf{p}}
\,u_{\mathbf{k}\mu}\left(\mathbf{r} \right) \right) \mathrm{.}
\end{equation}
Here $V$ is the volume of the simulation box and the functions
$u_{\mathbf{k},\nu}^\ast\left(\mathbf{r} \right)$ and
$u_{\mathbf{k},\mu}\left(\mathbf{r} \right)$ are the Bloch factors. Because of
the first Kronecker symbol the matrix elements are diagonal concerning the wave
number, which eliminates the $\mathbf{p}$ sum. The spin summation leads to an
additional factor of two. After using a relation between Fermi functions we
finally arrive at the following expression for the Onsager coefficients:
\begin{eqnarray}
\label{eq:kg-formula}
L_{mn}\left(\omega\right)&&=\frac{2\pi q^{4-m-n}}{3Vm_e^2\omega}
\sum_{\mathbf{k}\nu\mu}
 \langle \mathbf{k}\nu
|\hat{\mathbf{p}}|
\mathbf{k}\mu \rangle\cdot\langle \mathbf{k}\mu |\hat{\mathbf{p}}|
\mathbf{k}\nu \rangle \,
\\
&&\times
\epsilon_{\mathbf{k}\nu\mathbf{k}\mu}^{m+n-2}
\left(f_{\mathbf{k}\nu} - f_{\mathbf{k}\mu}\right)
\delta\left(E_{\mathbf{k}\mu}-E_{\mathbf{k}\nu}-\hbar\omega\right)
\mathrm{.} \nonumber
\end{eqnarray}

The Onsager coefficients (\ref{eq:kg-formula}) obey the symmetry relations
$L_{mn}(\omega)=L_{mn}(-\omega)$ and $L_{mn}(\omega)=L_{nm}(\omega)$. The
coefficient $L_{11}$ is known as the frequency dependent
Kubo-Greenwood-formula~\cite{Kubo1957,Greenwood1958} and has been widely applied
in FT-DFT-MD simulations.~\cite{Desjarlais2002,Clerouin2005,Kietzmann2007}

Similar but not identical frequency dependent formulas for $L_{12}$ and $L_{22}$
were given by Recoules {\it et al.}~\cite{Recoules2005} but not formally derived
in their work. The different formulations lead to deviations in the results for
the thermopower $\alpha$ and in the thermal conductivity $\lambda$ at nonzero
frequencies. Furthermore, the heat current from Ref.~\onlinecite{Recoules2005}
contains the chemical potential $\mu_e$ instead of the enthalpy $h_e$, see Eq.\ 
(\ref{ws}). However, it can be shown that the term proportional to the enthalpy
per particle $h_e$ has no influence on the thermal conductivity $\lambda$ in
one-component systems. As a consequence the numerical results of
Refs.~\onlinecite{Recoules2005, Recoules2009} could be reproduced by
Eq.\ (\ref{eq:kg-formula}) in the limit of $\omega\to0$ and are therefore not
questioned by the current work. Differences, however, occur in the static
results for the thermopower $\alpha$. 

The derivation in this chapter can be easily generalized to the spin dependent
form as well as to anisotropic systems.

\section{Results}\label{sec:res}

\subsection{Electrical conductivity}

\begin{figure}
\centering
\includegraphics[width=1.0\columnwidth]{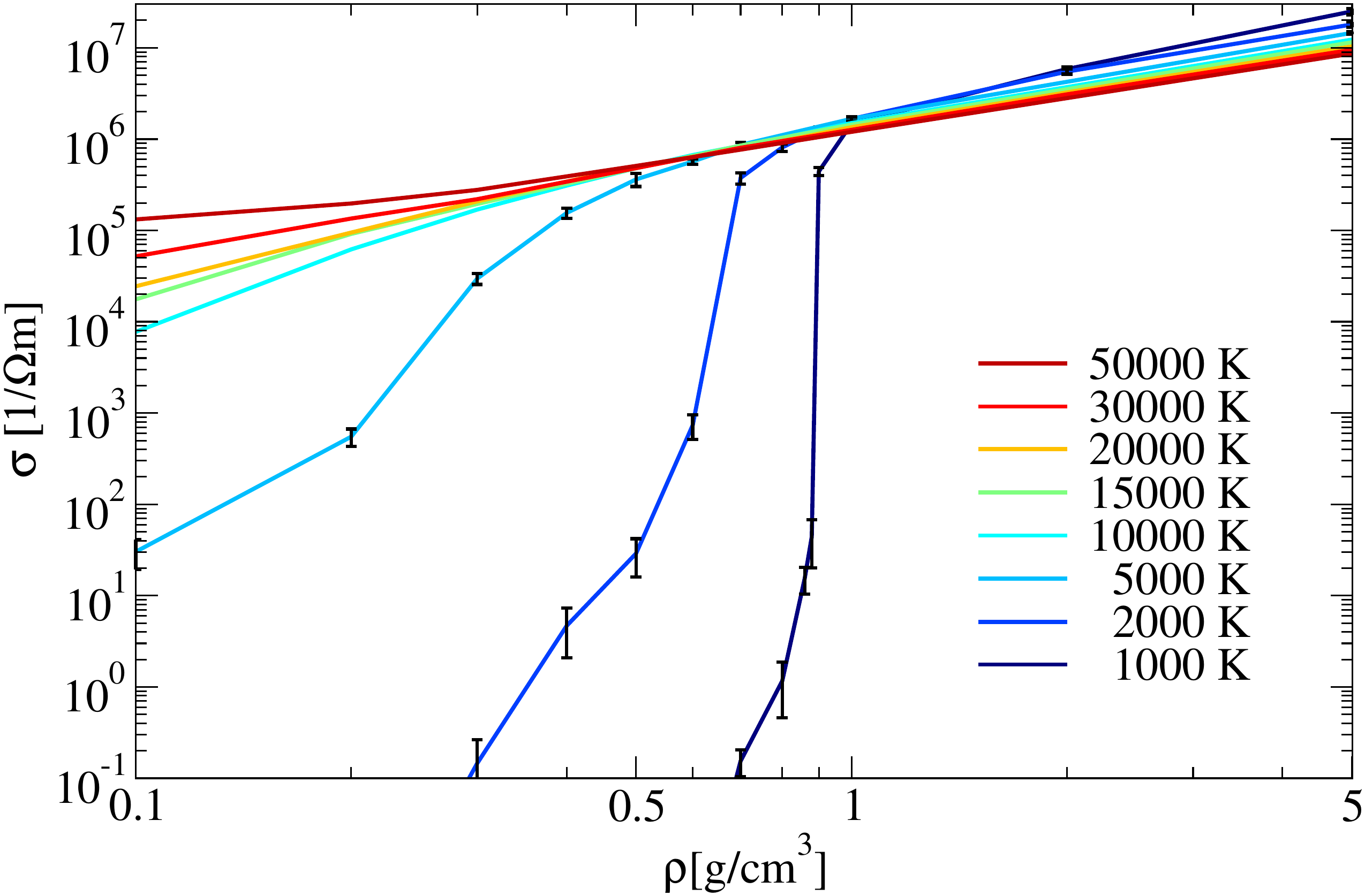}
\caption{(Color online) Electrical conductivity in dense hydrogen for different 
temperatures: from the region of the liquid-liquid phase transition up to the 
dense plasma.}
\label{fig:sigma}
\end{figure}

Fig.~\ref{fig:sigma} shows isotherms of the electrical conductivity over a wide
range of densities and temperatures. In previous work~\cite{Holst2008} it could
already be shown that these theoretical results agree well with data for the 
reflectivity and conductivity derived from shock-wave experiments. Here we 
concentrate on the general behavior of the theoretical curves on a large density 
and temperature scale and give results especially in the vicinity of the 
nonmetal-to-metal transition.

In general the conductivity rises in the whole range studied here with increasing 
density. On the one hand this is the result of pressure ionization which increases 
the amount of conducting electrons and on the other hand, even in the case of a
fully ionized system, a rising density of electrons results in a further increase 
of the conductivity.

At $T=1000$~K the liquid-liquid phase transition which was previously reported
for dense hydrogen~\cite{Morales2010a, Lorenzen2010} can be identified by the
steep increase of the conductivity over several orders of magnitude  
in a small density interval between $(0.7-0.9)$~g/cm$^3$. At temperatures 
higher than 1500~K the transition is continuous and the increase spreads to a 
larger range in density. At densities below this transition the electrical 
conductivity increases with temperature which is caused by thermal ionization 
of neutral particles. The additional free charges contribute to the conductivity. 

At densities above this transition the dependence on temperature is inverted: 
the electrical conductivity decreases with temperature, which is typical for 
metals. Increasing temperature broadens the Fermi function and therefore allows
additional electron scattering processes which reduce their mobility. As a result 
the conductivity is lower with increasing temperature. 

\begin{figure}
\centering
\includegraphics[width=1.0\columnwidth]{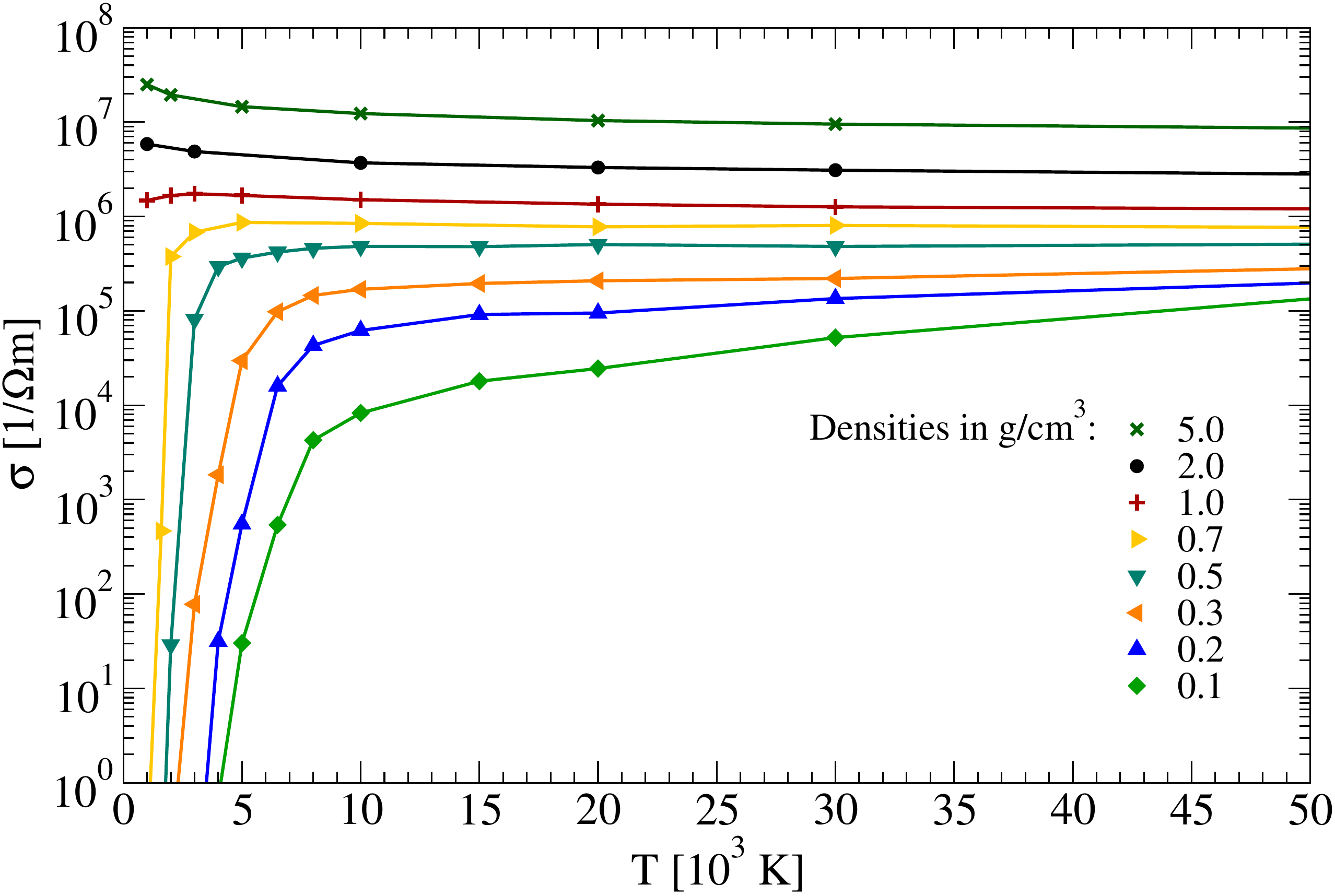}
\caption{(Color online) Electrical conductivity for different densities versus 
temperature.} \label{fig:sigmaic}
\end{figure}

This behavior is also illustrated in Fig.~\ref{fig:sigmaic} which shows the 
electrical conductivity along isochores for different temperatures. The 
conductivity decreases for temperatures above 2000~K along the isochores 
for densities higher than 0.9~g/cm$^3$ which are characteristic of the metallic 
phase. This indicates that most of the system is ionized and thus acts metal-like. 
Looking at lower densities the conductivity rises along the whole temperature 
range which is due to thermal ionization. The isochores clearly show the general 
behavior of a rising conductivity with increasing density.

\begin{figure}
\centering
\includegraphics[width=1.0\columnwidth]{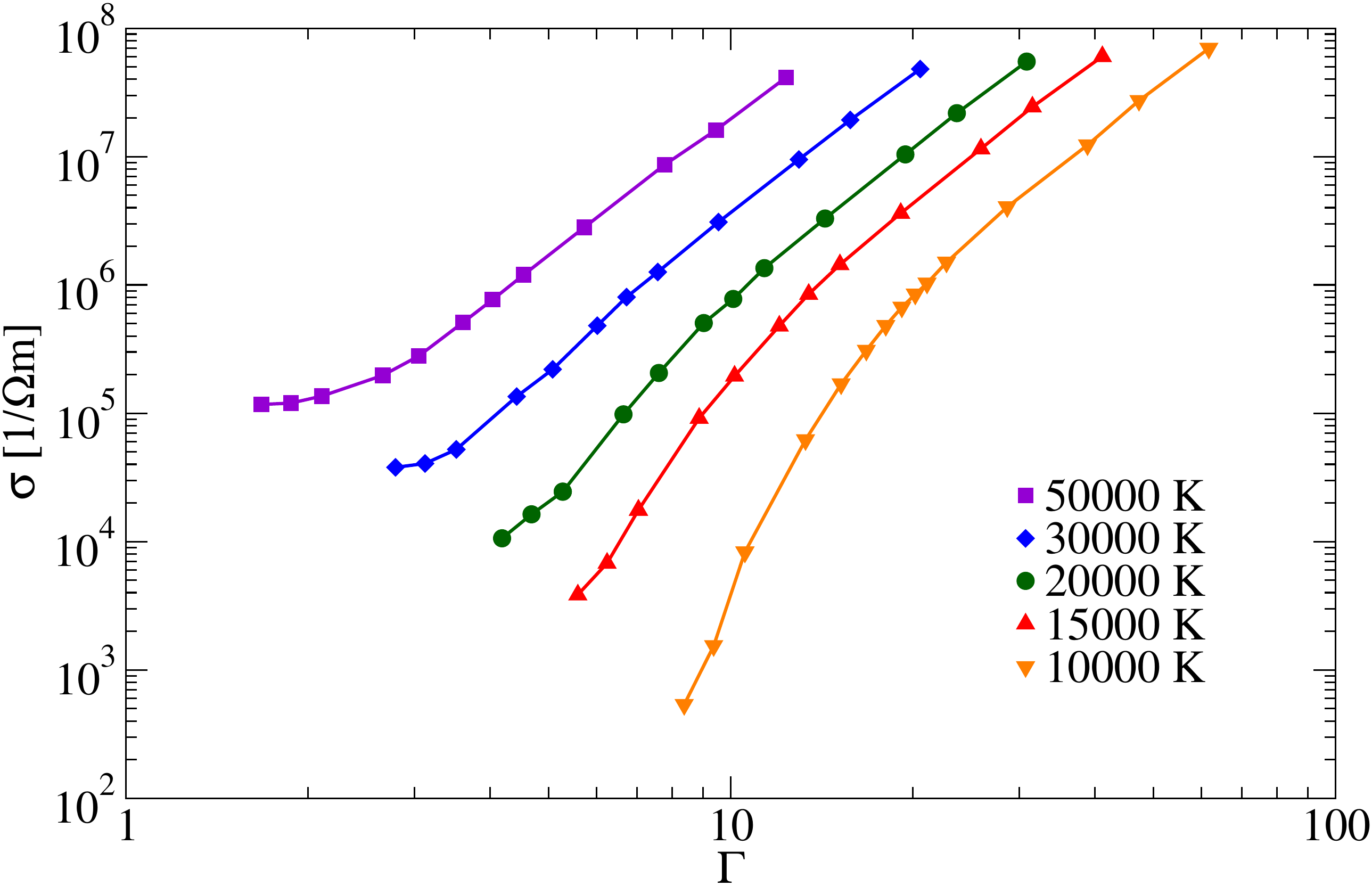}
\caption{(Color online) Electrical conductivity as function of the coupling
parameter $\Gamma$ for different temperatures. The plotted data cover a density
range of $(0.05-20)$~g/cm$^3$.}
\label{fig:siggam}
\end{figure}

Experiments in copper plasmas~\cite{Desilva1994} have indicated that the electrical
conductivity becomes a function of only the coupling parameter $\Gamma$ for values 
of $\Gamma\ge 10$. The plasma parameter $\Gamma$ is defined by
\begin{equation}\label{eq:Gamma}
\Gamma = \frac{e^2}{4\pi\varepsilon_0k_BT}\left(\frac{4\pi n_e}{3}
\right)^{\frac{1}{3}} \;\mathrm{,} 
\end{equation}
where $n_e$ is the number density of free electrons. Although such a behavior could 
not be confirmed later,~\cite{Desilva1998} a simple functional form of the electrical 
conductivity at high values of $\Gamma$ is still under discussion. To investigate 
whether or not such a simple scaling is valid in dense liquid hydrogen, the results 
for the electrical conductivity shown in Figs.\ \ref{fig:sigma} and \ref{fig:sigmaic} 
are plotted against $\Gamma$ in Fig.\ \ref{fig:siggam}. Only for temperatures
higher than 10000~K the system is strongly ionized. At lower
temperatures the occurrence of partial ionization prevents a proper calculation
of $\Gamma$ within FT-DFT-MD, because the method does not distinguish between bound 
and free electrons. Therefore we do not plot results for lower temperatures in Fig.\ 
\ref{fig:siggam}.

The isotherms appear to be almost parallel in this logarithmic plot and are clearly 
separated. Even at the highest available values for $\Gamma\le 60$ the isotherms do 
not tend to merge. We conclude that it is not possible to derive a simple 
temperature-independent relation for the conductivity that depends solely on the 
parameter $\Gamma$ as it was proposed earlier for other metallic liquids. 

\subsection{Thermal conductivity}

\begin{figure}[htb]
\centering
\includegraphics[width=1.0\columnwidth]{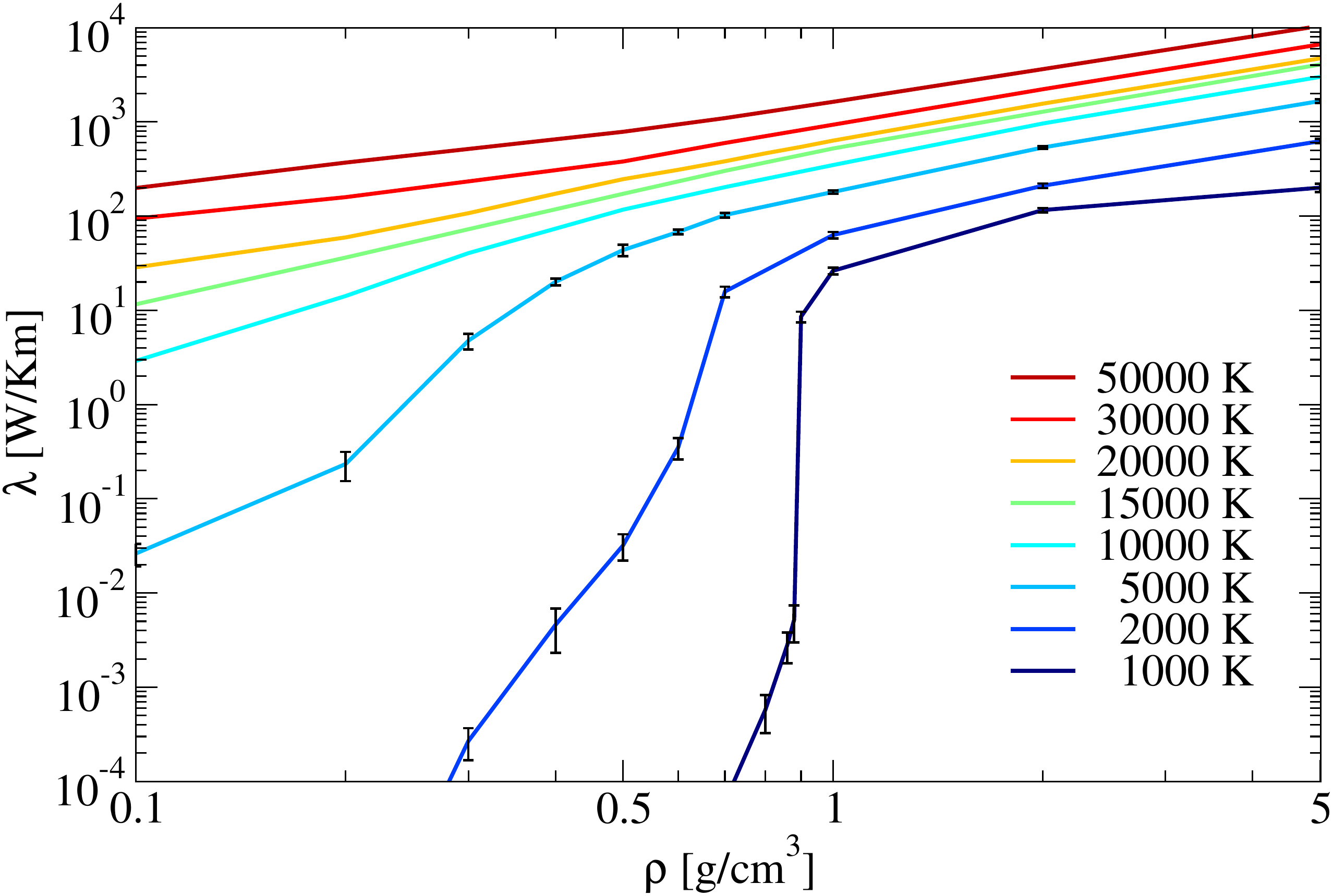}
\caption{(Color online) Thermal conductivity versus density for different
temperatures.} \label{fig:lambda}
\end{figure}

The isotherms of the thermal conductivity are plotted in Fig.\ \ref{fig:lambda}. 
They show a similar behavior as the electrical conductivity and indicate a sharp
nonmetal-to-metal transition at a temperature below 1500~K. At this transition
the thermal conductivity increases over several orders of magnitude in a narrow 
density range at about 0.9~g/cm$^3$. The increase in the thermal conductivity is 
due to the growing number of delocalized electrons which are produced along the 
phase transition. Above the critical temperature this transition becomes broader 
and is caused by a combination of pressure and temperature ionization. 

In contrast to the behavior of the electrical conductivity, the thermal
conductivity does not decrease with temperature in the metallic phase. The
isotherms increase systematically with temperature for all densities.

\begin{figure}[htb]
\centering
\includegraphics[width=1.0\columnwidth]{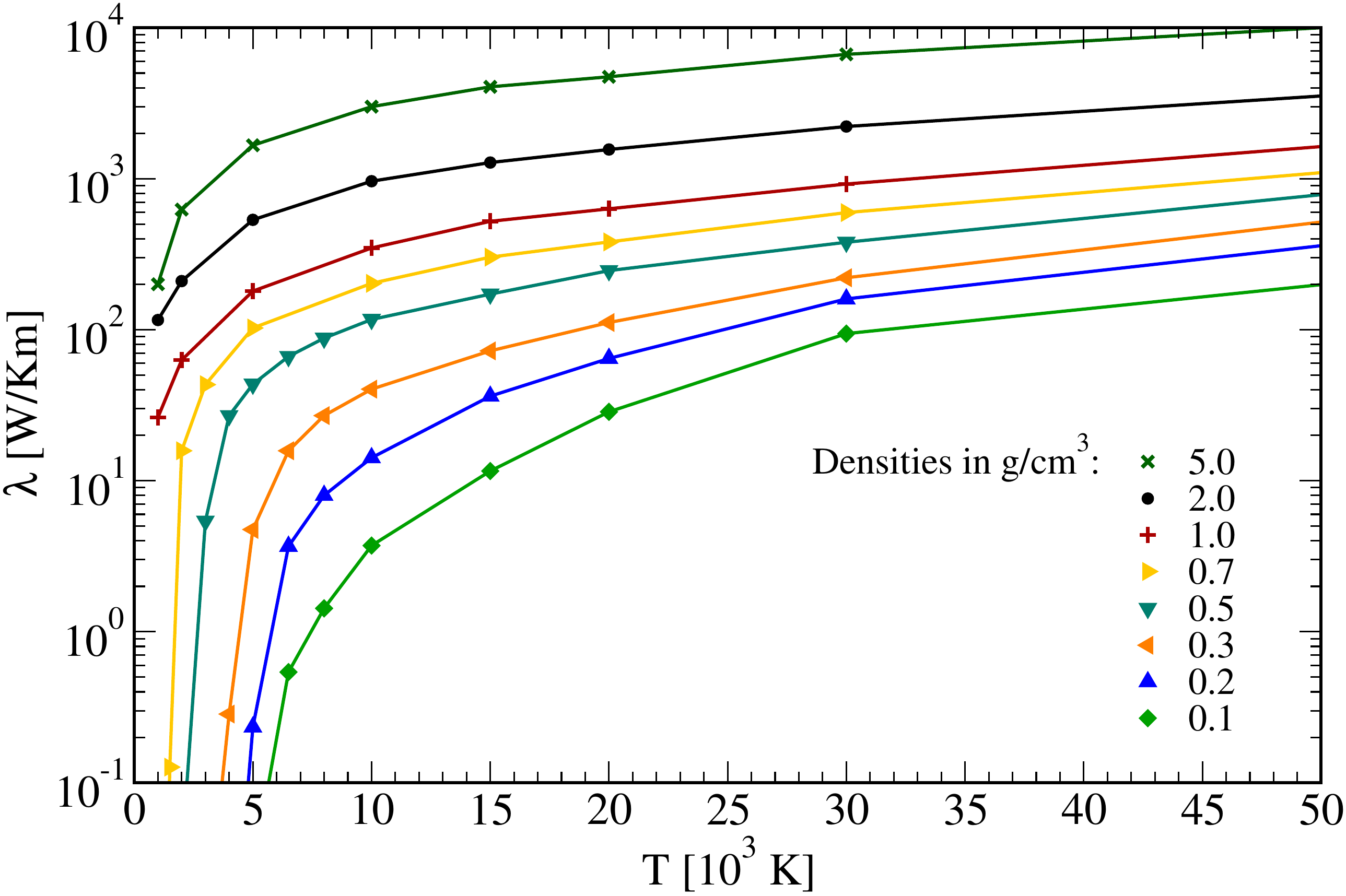}
\caption{(Color online) Thermal conductivity versus temperature for different
densities.} \label{fig:lambdaic}
\end{figure}

This is also shown along the isochores of the thermal conductivity that are
plotted in Fig.~\ref{fig:lambdaic}. These curves depict likewise that the
thermal conductivity rises invariably with increasing density and temperature.

\subsection{Thermopower}

The thermopower $\alpha$ characterizes the generation of an electric field as
response to a temperature gradient. For most systems this electric field has a 
direction opposite to the temperature gradient which results in a negative
thermopower. The thermopower is most sensitive to changes in the electronic 
structure since it can be expressed as the derivative of the logarithm of the 
electronic conductivity with respect to the energy at the Fermi 
surface.~\cite{MottDavis1979} Such a relation which is also known as Mott 
formula follows from the Kubo-Greenwood equation (\ref{eq:kg-formula}) in the 
degenerate domain under strong scattering conditions. 

Interestingly, large positive values for the thermopower were measured in fluid
mercury~\cite{Gotzlaff1988, Hensel1999} near the liquid-vapor critical point
which is located at $T_c=1751$~K and $\varrho_c=5.8$~g/cm$^3$. In this region,
isotherms of the electrical conductivity near $T_c$ show a strong increase with
the density which is steepest just at the critical density $\varrho_c$. This
behavior was assumed to be related to fluctuations in the electron density which
are pronounced near the critical point due to critical fluctuations. In
particular, a zero of the thermopower was observed exactly at the critical
density. The interesting question arises whether or not a zero of the
thermopower is a precursor of a first-order phase transition in dense liquids
which undergo a nonmetal-to-metal transition. Wide regions with a positive
thermopower have been predicted for dense hydrogen by a simple chemical
model~\cite{Hoehne1984} but were not confirmed in an advanced chemical
approach~\cite{Reinholz1995} by {\it ab initio} simulations or by experiments
yet.

\begin{figure}[htb]
\centering
\includegraphics[width=1.0\columnwidth]{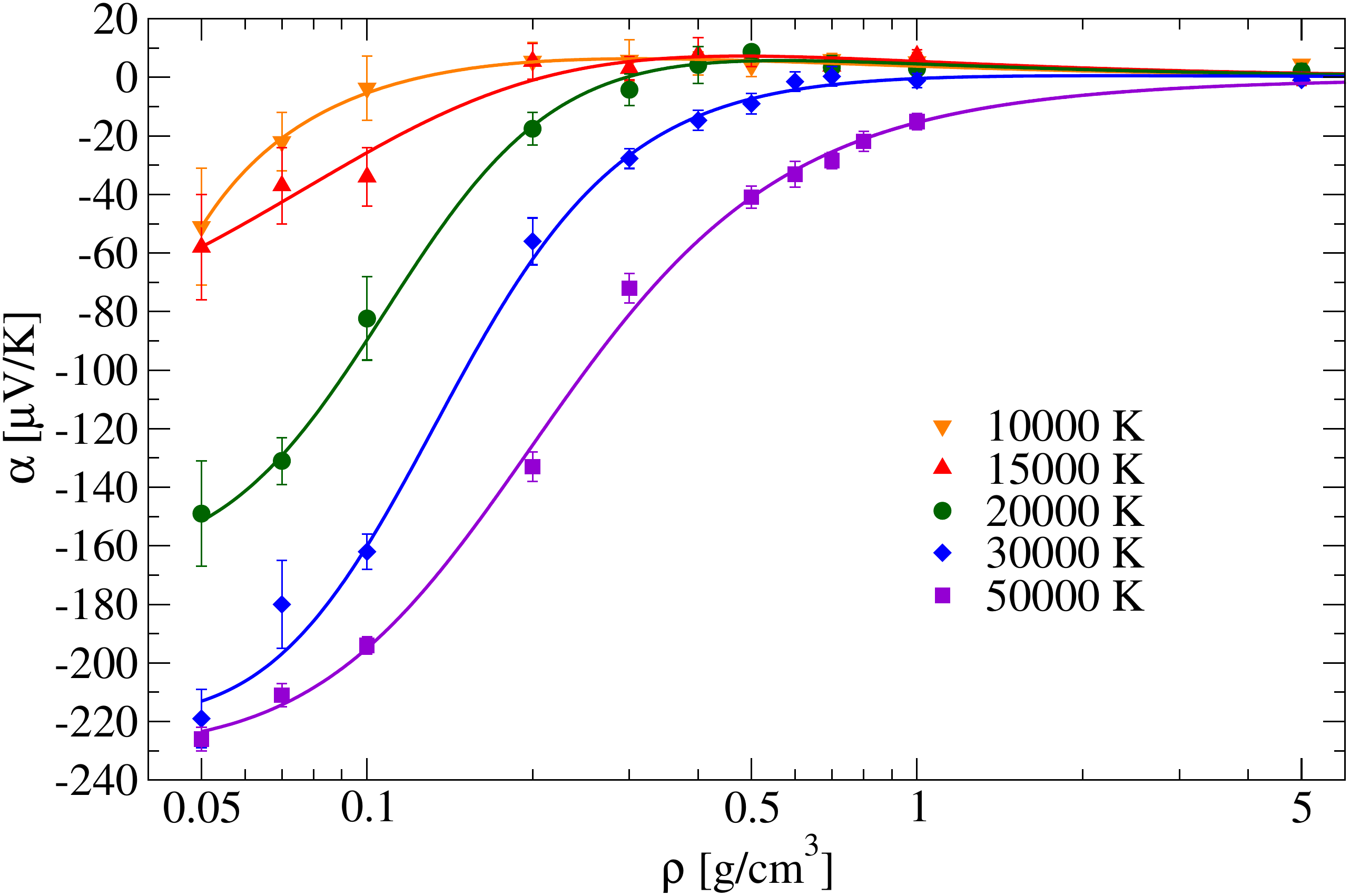}
\caption{(Color online) Thermopower versus density for different temperatures.}
\label{fig:alpha}
\end{figure}

In Fig.~\ref{fig:alpha} isotherms of the thermopower $\alpha$ are plotted as
function of the density. The symbols represent the results from the simulations
and the error bars show the statistical uncertainties. As guide to the eye,
polynomial functions were fitted to the numerical results. The thermopower is
mostly negative and reaches a value of about zero at high densities. At lower
densities the thermopower decreases. The low density limit of
$\alpha=-60.60$~$\mu$V/K is known from the Spitzer theory, see e.g.
Ref.~\onlinecite{Reinholz1995}. With higher temperatures the negative values
become systematically larger. We expect that these values become smaller again
at low densities to reach the Spitzer limit. The thermopower shows positive mean
values below 20000~K and between 0.2~g/cm$^3$ and 0.5~g/cm$^3$. These values are
very small and, in fact, about the size of the uncertainties of the respective
mean values. We therefore cannot conclude definitely that significantly positive
values occur within the accuracy of our {\it ab initio} calculations. To answer
the question whether or not the zero of the thermopower occurs at the critical
point of the liquid-liquid phase transition in dense hydrogen which is
predicted~\cite{Morales2010a,Lorenzen2010} at about $0.8$~g/cm$^3$, the
thermopower has to be evaluated for temperatures below 2000~K since
$T_c=(1400-2000)$~K. For this region, reasonable results can be given for the
electrical and thermal conductivity but not for the thermopower. This is because
the values of the Onsager coefficients each decrease over orders of magnitude in
systems with a majority of localized electrons while especially the decrease of
$L_{12}$ is two orders of magnitude weaker than that of $L_{11}$. This behavior
on the one hand causes the second contribution to the thermal conductivity in
Eq.~(\ref{eq:trans_coeff}) to vanish, because $L_{12}$ is squared in the
numerator. As a result the thermal conductivity becomes proportional to $L_{22}$
which can be evaluated successfully in this case, as well as $L_{11}$. On the
other hand this leads to an enormous increase in the statistical fluctuations of
the thermopower, which causes uncertainties that reach hundreds of $\mu$V/K.

\begin{figure}[htb]
\centering
\includegraphics[width=1.0\columnwidth]{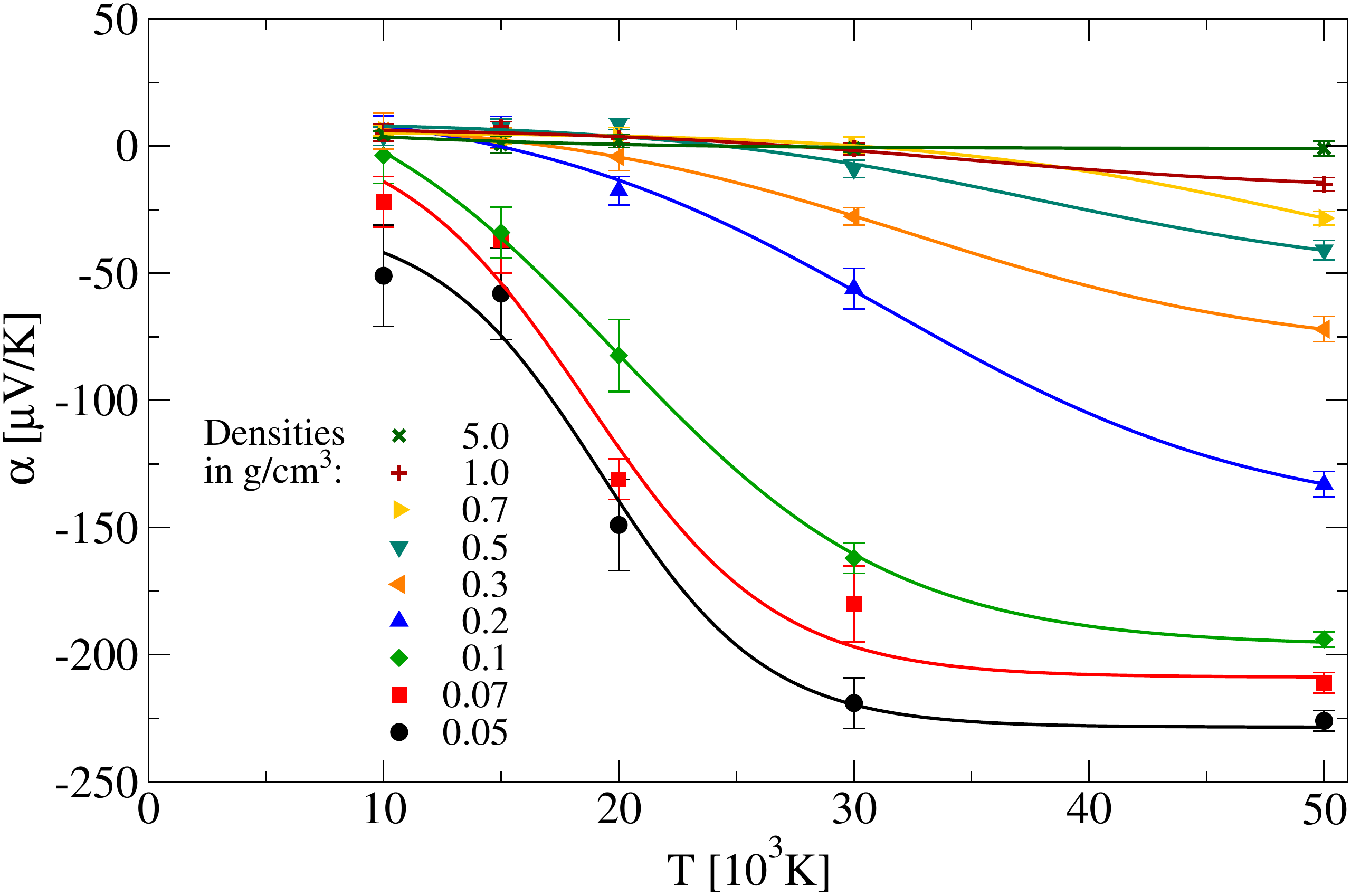}
\caption{(Color online) Thermopower versus temperature for different densities.}
\label{fig:alphaic}
\end{figure}

In Fig.~\ref{fig:alphaic} isochores of the thermopower $\alpha$ are plotted.
The thermopower decreases with higher temperature, which is more pronounced the
lower the density is. Positive values appear at low temperatures, but the
statistical error is too large to prove this result unambiguously	.

\subsection{Lorenz number}

The Lorenz number Eq.\ (\ref{LorenzN}) describes mainly the relation between
thermal and electrical conductivity. For simple metals this relation is
described via a constant and is known as the Wiedemann-Franz law. For high
degeneracy this constant is $L_0=\pi^2/3$.~\cite{Ashcroft1976} This relation can
be used to easily obtain the thermal conductivity for metals if the electrical
conductivity is known. Here we calculate the Lorenz number for dense hydrogen in
order to identify the region where the Wiedemann-Franz law is valid. High
degeneracy occurs only at sufficiently high densities so that we expect
deviations from the Wiedemann-Franz law at lower densities and higher
temperatures. 

\begin{figure}[htb]
\centering
\includegraphics[width=1.0\columnwidth]{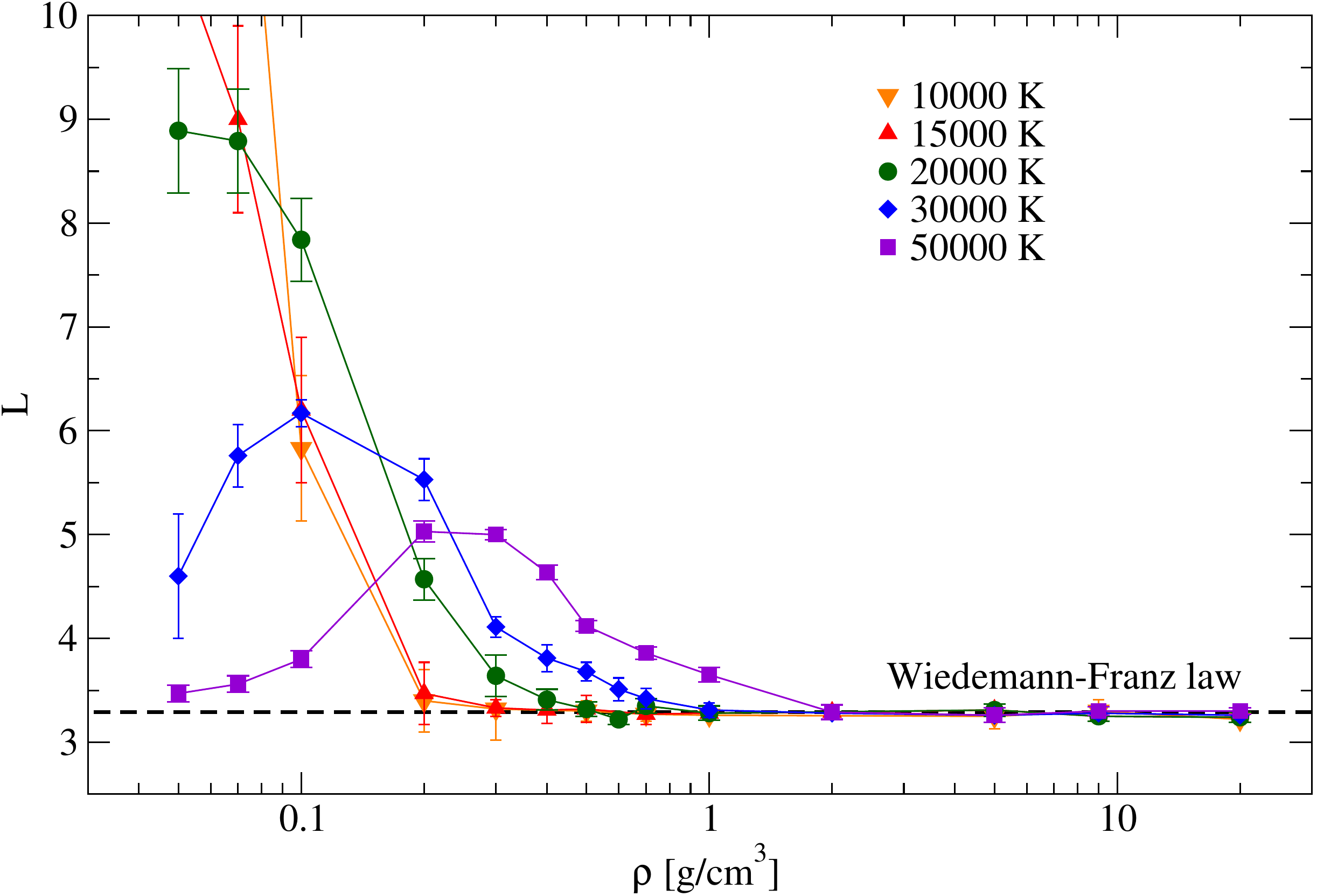}
\caption{(Color online) Lorenz number versus density for different temperatures.
The dashed black line displays the limiting value of $L_0=\pi^2/3$.}
\label{fig:lorenz}
\end{figure}

The Lorenz number is shown in Fig.~\ref{fig:lorenz} for several temperatures. At
densities above 1~g/cm$^3$ the Lorenz number is almost constant and shows no
temperature dependence which indicates that the Wiedemann-Franz law is valid
there. The value of $L_0=\pi^2/3$, which is shown by a dashed black line, could
be reproduced for each temperature at densities higher than (1-2)~g/cm$^3$
within the statistical uncertainties. This behavior is consistent with the
metallic-like properties observed already for the conductivity at high
densities.

At densities below (1-2)~g/cm$^3$ the Lorenz number shows strong deviations from 
the Wiedemann-Franz law with a pronounced temperature dependence. In particular, 
the Lorenz number rises strongly with decreasing density. 
This behavior becomes more pronounced at lower temperatures. On the other hand, 
at the lower temperatures the validity of the Wiedemann-Franz law extends to 
smaller densities, e.g.\ down to 0.2~g/cm$^3$ at 10000~K and 15000~K.  

Along the isotherms of 30000~K and 50000~K a maximum can be identified. Thus, 
at lower densities the Lorenz number decreases again. At 30000~K this maximum 
is found at a smaller density than at 50000~K and shows a larger value. This
indicates that with decreasing temperature the maximum shifts systematically to 
smaller densities and increases. For temperatures below 30000~K the maximum is 
not displayed. It appears to be situated at densities below 0.05~g/cm$^3$, which 
is the minimal density calculated here and which displays the current limit of 
our computational capabilities. However, we expect a maximum to appear for all 
temperatures because the Lorenz number is also known exactly at low densities 
from the Spitzer theory for fully ionized plasmas.~\cite{Spitzer1953} The latter 
predicts a constant low-density value of 1.5966, see e.g.\
Ref.~\onlinecite{Reinholz1995}.

Our results indicate that the deviations from the Wiedemann-Franz law might
reach a full order of magnitude in certain regions of density and temperature.
Further research will be necessary to investigate this important aspect in more
detail. We also expect that certain deviations from the Wiedemann-Franz law
occur in warm dense matter of arbitrary composition. In hydrogen, we predict 
that the Wiedemann-Franz is valid at densities above 2~g/cm$^3$ for temperatures 
below 50000~K, see Fig.~\ref{fig:lorenz}.

\section{Conclusions}\label{sec:concl}

We derived formulas for the Onsager transport coefficients $L_{mn}$ within the
Kubo theory for application in {\it ab initio} simulations. Using these
expressions we calculated the electrical and thermal conductivity as well as
the thermopower of liquid hydrogen for a wide range of temperatures and densities 
in the megabar range. In particular, we characterize the nonmetal-to-metal
transition in hydrogen by observing a rapid increase in both the electrical and
the thermal conductivity. The thermopower shows a trend towards positive values 
in a region where the critical point of the liquid-liquid phase transition is 
expected, similar to the behavior of liquid mercury. At low temperatures more
accurate calculations will be necessary for this transport coefficient which are
beyond the scope of the currently available computer capacity. The {\it ab
initio} calculation of the Lorenz number shows in addition that the validity of
the original Wiedemann-Franz law is limited to the metallic-like regime of
hydrogen. The theoretical framework given here can be applied to bulk material
calculations for arbitrary materials within FT-DFT-MD.

\begin{acknowledgments}

We thank Michael P.\ Desjarlais, Winfried Lorenzen, Thomas R.\ Mattsson, Stephane
Mazevet, and Vanina Recoules for many helpful discussions. This work was
supported by the Deutsche Forschungsgemeinschaft within the SFB~652 and the
DFG project SPP~1488 (planetary magnetism), and the North-German
Supercomputing Alliance (HLRN).
\end{acknowledgments}

\bibliography{lik}

\begin{thebibliography}{73}
\bibitem [{VAS()}]{VASPOPTIC}%
  \BibitemOpen
  \href@noop {} {}\bibinfo {note} {The matrix elements are extracted out of the
  OPTIC file. It has to be ensured that all Bloch states are considered as
  initial and final.}\BibitemShut {Stop}%
\end{thebibliography}

\end{document}